\documentstyle[epsf]{mn}

\def\bxi{{{\bar \xi}}}
\def\ltsima{$\; \buildrel < \over \sim \;$}
\def\lsim{\lower.5ex\hbox{\ltsima}}
\def\gtsima{$\; \buildrel > \over \sim \;$}
\def\gsim{\lower.5ex\hbox{\gtsima}}

\title[Does Gravitational Clustering Stabilize On Small Scales?]
{Does Gravitational Clustering Stabilize On Small Scales?}

\author[Bhuvnesh Jain]
	{Bhuvnesh Jain \\
	Max-Planck-Institut f\"ur Astrophysik,
Karl-Schwarzschild-Strasse 1, 85740 Garching, Germany\\
\smallskip 
Email: bjain@mpa-garching.mpg.de\\
}
\date{}

\begin{document}

\maketitle

\begin{abstract}

The stable clustering hypothesis is a key analytical anchor on the 
nonlinear dynamics of gravitational clustering in cosmology. It states
that on sufficiently small scales the mean pair velocity approaches
zero, or equivalently, that the mean number of neighbours of a particle
remains constant in time at a given physical separation. N-body 
simulations have only recently achieved sufficient resolution to probe
the regime of correlation function amplitudes $\xi \sim 100-10^4 $ in
which stable clustering might be valid. In this paper we use N-body 
simulations of scale free spectra $P(k)\propto k^n$ with $-2\leq n
\leq 0$ and of the CDM spectrum to apply two tests for stable
clustering: the time evolution and shape of $\xi(x,t)$, and the mean 
pair velocity on small scales. We solve the pair conservation equation
to measure the mean pair velocity, as it provides a more accurate
estimate from the simulation data. For all spectra the results are 
consistent with the stable clustering predictions on the smallest
scales probed, $x < 0.07 \ x_{nl}(t)$, where $x_{nl}(t)$ is the
correlation length. The measured stable clustering regime corresponds 
to a typical range of $200\lsim \xi \lsim 2000$, though spectra with 
more small scale power ($n\simeq 0$) approach the stable clustering 
asymptote at larger values of $\xi$. 

We test the amplitude of $\xi$ predicted by the
analytical model of Sheth \& Jain (1996), and find agreement to within
$20\%$ in the stable clustering regime for nearly all spectra. For the
CDM spectrum the nonlinear $\xi$ is accurately approximated by this
model with $n \simeq -2$ on physical scales  $\lsim 100-300 h^{-1} 
{\rm kpc}$ for $\sigma_8=0.5-1$, and on smaller scales at earlier times. 
The growth of $\xi$ for CDM-like models is discussed in the context of
a power law parameterization often used to describe galaxy 
clustering at high redshifts. The growth parameter $\epsilon$ is
computed as a function of time and length scale, and found to be
larger than $1$ in the moderately nonlinear regime -- thus the growth
of $\xi$ is much faster on scales of interest than is commonly assumed. 

\end{abstract} 


\maketitle

\section{Introduction}
The stable clustering hypothesis is one of the few analytical handles
on the deeply nonlinear regime of gravitational clustering. 
It states that the mean relative velocity of particle pairs
in physical coordinates is zero. Hence in 
comoving coordinates the mean relative velocity
exactly cancels the Hubble recession velocity between pairs of
particles. As we will see below, this is 
equivalent to the statement that the mean number of neighbours
of a particle in physical coordinates remains constant in time. 
The stable clustering hypothesis invokes the physical 
picture of a virialized cluster which has separated out from the
expanding background, and is neither expanding nor 
contracting. Since on small scales, any statistical measure is 
dominated by the contribution from dense clusters, the clustering
should statistically be stable. 

This hypothesis leads to predictions for the
evolution in time of the autocorrelation function $\xi$. In the case of 
scale free spectra which display self-similar scaling, the slope of $\xi(x)$
in the small scale regime is predicted as well. For $\xi\gg 1$ it is
extremely difficult to make analytical approximations to the growth 
of clustering. Hence the stable clustering hypothesis has been very
useful in relating the shape of $\xi$ and the power spectrum $P(k)$
in the nonlinear regime to the initial spectrum. Peebles (1974) and
Gott \& Rees (1975) implemented the first such applications. 
The widely used property of hierarchical scaling of the higher order
moments of the density in terms of its second moment also derives from
the dynamics of stable clustering. As emphasized in the next section, 
this property requires additional assumptions of stability at each
order in the distribution. 

N-body simulations are the ideal tool to test stable clustering. 
Efstathiou et al. (1988) carried out a pioneering study of
self-similar evolution, and tested the small scale $\xi$ and $v$
for stable clustering. Their data were consistent with 
the stable clustering slope for $\xi$ on the smallest scales, but
their $32^3$ particle simulations lacked the resolution to give
any definitive conclusion. 
Recently Padmanabhan et al. (1995) and Colombi, Bouchet \& Hernquist
(1995; hereafter CBH), have tested the stable clustering predictions
for $\xi, v$, and in the latter case for the hierarchical form of
the higher moments $S_3$, $S_4$ and $S_5$ as well. While Padmanabhan
et al. (1995) found departures from stable clustering in their
$\Omega=1$ simulations, CBH verified the stable
clustering prediction in their tests of $\xi$. However, they
found small departures from the hierarchical relation for the higher
moments. 

In this paper we use ${\rm P^3 M}$ (particle-particle/particle-mesh) 
simulations with $100^3-144^3$ particles 
to test stable clustering for power law spectra with 
$-2\leq n\leq 0$, and for the CDM spectrum. We measure $\xi$ and 
$v$, for which we use the pair conservation equation to obtain
estimates with greater accuracy than allowed by direct measurements. 
Our approach and numerical
resolution is similar to that of CBH, with the advantage that 
we have $4-10$ times as many particles. We do not however test for
the higher moments as they do. While stable clustering predicts the
slope of $\xi$, its amplitude can be approximated by additional 
assumptions as done in the analytical model of Sheth \& Jain (1996). 
We test their predictions against the measured amplitude in the N-body
data. We also test for stable clustering in two high resolution
CDM simulations and identify the range of scales and epochs on which 
it is an adequate approximation. 

We begin in the next section with the relevant BBGKY equations 
through which the stable clustering hypothesis leads to 
the predictions for $\xi$ and the higher moments. In Section 3 we
outline a secondary infall model which can be connected with the
nonlinear form of $\xi$, and discuss the dynamical effects which
might invalidate stable clustering. Section 4 contains 
a description of the N-body simulations, and an analysis
of the effects of limited numerical resolution on small scales. 
Section 5 presents the main results for $\xi$, with the mean pair velocity
results in Section 5.1, and results for the CDM spectrum in Section 5.2.
We conclude in Section 6. 

\section{The BBGKY hierarchy equations}

The stable clustering hypothesis yields the form of the nonlinear 
correlation function when applied to the moments of the BBGKY hierarchy
equations. The presentation below follows that of Peebles (1980, Sections
71 and 73). 

The second equation of the BBGKY hierarchy of equations,
when integrated over momenta, yields the pair conservation equation. 
This equation connects the rate of change of the autocorrelation 
function $\xi(x,t)$ to the mean pair velocity in comoving coordinates, 
$v(x,t)=a\dot x$:
\begin{equation}
{\partial \xi \over \partial t} + {1\over x^2 a}{\partial\over
\partial x}\left[ x^2(1+\xi)v\right]=0 .
\label{pair}
\end{equation}
Integrating the above equation over a sphere of radius $x$ gives
\begin{equation}
{\partial \over \partial t}\left[na^3 4\pi\int_0^{x}dx' x'^2 
(1+\xi) \right]
+4\pi a^2 x^2 n (1+\xi) \, v =0 \, ,
\label{pairint}
\end{equation}
where $n(t)\propto a^{-3}$ is the number density in physical
coordinates. 
From the definition of $\xi$ it follows that
$n(t)a^3 4\pi x^2 dx [1+\xi(x,t)]$ is the typical number of neighbours 
in a spherical shell of thickness $dx$ at a 
distance $x$ from a particle. Thus the first term in equation
(\ref{pairint}) is the rate of change of neighbours inside a sphere
of radius $x$, while the second term is the flux of particles through
the surface of the sphere -- hence the name pair conservation 
equation. 

The stable clustering hypothesis is the statement that the relative 
pair velocity in comoving coordinates, $a\dot x$, 
exactly cancels the Hubble 
velocity, $\dot a x$, so that the mean pair velocity in physical 
coordinates $d(ax)/dt=0$. On substituting this assumption, 
\begin{equation}
v(x,t)=-\dot a x ,
\label{stable1}
\end{equation}
in (\ref{pair}), the functional form of $\xi(x,t)$ which satisfies 
the equation becomes restricted to
\begin{equation}
1+\xi= a^3 f\left[a(t)x\right] ,
\label{stable2}
\end{equation}
and we can approximate $1+\xi\simeq \xi$, as the stable 
clustering hypothesis is
applied on very small scales were $\xi$ is at least of order
$100$. Equation (\ref{stable2}) thus fixes the growth of $\xi$
in physical coordinates: $\xi(r, t)\propto a^3$. 
Physically it means that the typical
number of neighbors per unit volume, at small enough separations
$r$ from a particle, remains 
constant in time, as it is $\simeq n(t) \xi(r, t)$, 
and $n(t)\propto a^{-3}$ as noted above. 

If the universe is spatially flat, with $\Omega=\Omega_{matter}=1$,
and the initial spectrum of perturbations is a pure power law 
$P(k)\propto k^n$, the subsequent evolution is expected to be 
self-similar (Peebles 1980, Section 73). Then
$\xi(x,t)$ and $P(k,t)$ take the special functional form:
\begin{equation}
\xi(x,t)=\hat\xi({x/x_0a^\alpha})\, \, ;
\, \, P(k,t)=a^{3\alpha}k_0^{-3}\, 
\hat P({k a^{\alpha}/ k_0})\, ,
\label{psim}\end{equation} 
where $\alpha=2/(3+n)$;
$k_0,\, x_0$ are constants which must be determined from the initial 
conditions; and $\hat P, \, \hat \xi$ are unspecified dimensionless
functions. It is easy to verify that the linear spectrum 
$P_1(k,t)\propto a^2 k^n$ is consistent with the functional
form of equation (\ref{psim}). The linear mean squared density contrast
on length scale $k^{-1}$ scales as $a^2 k^{-(3+n)}$, hence 
equating it to unity gives the same scaling, $k\propto a^{-2/(3+n)}$
as contained in equation (\ref{psim}). This unique
scaling in time of characteristic length scales is the essence of
self-similar evolution. 

Equating the functional forms of $\xi$ given by 
equations (\ref{stable2}) and (\ref{psim}) restricts the
$t$ and $x$ dependence of $\xi$ to be a pure power law. The solution
for $\xi$, and similarly for $P$, is:
\begin{equation}
\xi(x,t)\propto (x/a^{\alpha})^{-\gamma} \, \, ;
\, \, P(k,t)\propto a^{6\over (5+n)} k^{-6\over (5+n)}\, ,
\label{stable3}\end{equation}
where $\gamma=(9+3n)/(5+n)$. 
Thus the stable clustering assumption, in conjunction with the
requirement of self-similarity, fixes both the asymptotic growth
and the shape of the power spectrum/correlation function. While
self-similarity is an idealization as it requires the initial spectrum
to be a pure power law, it can be a useful
approximation to a more realistic description, and as we shall see
it is very useful for testing the dynamical validity of 
stable clustering. 

For completeness we re-write the pair conservation equation
(\ref{pairint}) in terms of the mean interior correlation function, 
which is the integral of the correlation function over a sphere,
divided by the volume of the sphere: 
\begin{equation} 
\bxi(x,t)\equiv {3\over x^3} \int_0^x dx \,  x^2 \, \xi(x,t)\, . 
\label{corrint}\end{equation}
Using this definition and re-arranging terms in equation
(\ref{pairint}) gives
\begin{equation}
{a\over 3(1+\xi)}{\partial \bxi\over \partial a} = {-v\over \dot a x}
\equiv {-v \over H r}\, ,
 \label{stable4}\end{equation}
where $H r$ is the Hubble velocity on the physical scale $r=ax$. 
In Section 4 we shall use this form of the pair conservation equation 
to measure the mean pair velocity $v$ from the N-body simulations. 

\subsection{Hierarchical relation of higher moments}
 
The hierarchical form of the 3-point function
$\zeta_{123}
\equiv\zeta(r_{12},r_{23},r_{31})$ and of the higher moments 
can also be related to the stability
assumption. Here we outline the argument which motivates the
hierarchical form using self-similarity and stable clustering. 
The purpose of 
this exercise is to point out that additional assumptions of
stability at every order in the particle distribution are needed
to obtain the hierarchical form at that order. These go beyond the 
assumption $-v/Hr=1$ used to obtain the stable clustering shape of $\xi$. 

Following Peebles (1980), consider the equation of
conservation of triplets of particles, obtained by integrating
over momenta the 3rd equation of the BBGKY hierarchy. In terms of
the relative velocities of particle pairs, it can be written as 
\begin{equation}
a \frac{\partial h_3}{\partial t} + \frac{\partial}{\partial \vec
x_{12}} \cdot \vec w_{12} h_3 + \frac{\partial}{\partial \vec
x_{31}} \cdot \vec w_{31} h_3 \, = \, 0\, ,
\label{triplet}\end{equation}
where $h_3=1+\xi_{12}+\xi_{23}+\xi_{31}+\zeta_{123}$, and $\vec w_{ij}$
is the mean pair velocity of particles $i,j$ given the position of the
third particle in the triangle formed by the triplet. This definition 
makes $\vec w_{12}$
a different statistic from $\vec v_{12}$: the former is a 3-point
statistic as it is obtained by integrating over momenta the
3-point distribution function, while the latter is a 2-point
statistic.  

In equation (\ref{triplet}) if we assume that $w_{12}=-\dot a x_{12}$
and $w_{31}=-\dot a x_{31}$
on small scales, then we are led to the functional form 
$h_3 \equiv a^6 h(ax_{12},ax_{23})$. Combined with the constraint of
self-similarity this leads to a power law solution for $h_3$ and 
therefore for the three point function: $\zeta \propto x^{-2 \gamma}$, 
where $x$ is the size of the triangle, and $\gamma$ is the stable
clustering slope of $\xi$ defined following equation (\ref{stable3}). 
It is thus consistent with the form 
\begin{equation}
\zeta_{123}\, =\, Q
\left(\xi_{12}\xi_{23}+\xi_{23}\xi_{31}+\xi_{31}\xi_{12}\right) \, ,
\label{hier}\end{equation}
which is the widely used hierarchical form for the 3-point
function. This argument can be generalized to the 4- and 5-point functions, 
and has been found to agree with results of N-body simulations as
well as observations of galaxies. 

The argument leading to the hierarchical form for $\zeta$
makes an assumption about the 3-point particle distribution.
Likewise for the $N$-th order correlation
function, the assumption required is that the mean relative 
velocity of particle pairs, given
the presence of $N-2$ other particles, be zero. Therefore 
additional conditions of stability are needed at each order of the
hierarchy to get the analogues of equation (\ref{hier}) for 
higher order moments. These conditions clearly go beyond the assumption 
$v=-\dot a x$ which led to 
equation  (\ref{stable3}) for the second moment $\xi$. 
The tests performed in this paper relate
to $\xi$, and therefore do not necessarily verify the assumptions 
leading to the hierarchical form of higher
moments. This is also relevant in comparing our results with 
those of CBH, who appear to find small departures from the
hierarchical relation for the $S_3, S_4$ and $S_5$ parameters
(while their results for $\xi$ are consistent with ours as
discussed in Section 5). 

\section{Analytical Models}

For convenience in modelling, small scale clustering 
can be thought of in two somewhat distinct ways. 
One is the hierarchical growth of structure by the continuous mergers
of smaller clumps. N-body simulations have highlighted the 
importance of mergers, but 
the detailed dynamics is very difficult to model analytically. The
other model, known as secondary infall, visualizes a 
more gradual and spherically symmetric accretion of matter 
on to initial density peaks leading to the formation of a halo. 
Simulations show that spherically
symmetric accretion rarely occurs in the formation of typical
halos. But one might imagine that the average properties of halos
are well approximated by the outcome of such a secondary infall
model. 

The advantage of the secondary infall picture is that it
is analytically tractable, and the density profile of a halo
resulting from a given initial profile can be calculated. 
Gunn \& Gott (1972) and Gott (1975) made the first calculations
of spherical accretion models, 
while Fillmore \& Goldreich (1984) and Bertschinger (1985)
obtained detailed solutions for the profile arising from
the late time behavior of particle orbits. 
Hoffman \& Shaham's (1987) influential paper
applied these results to the structure of halos formed in a cosmological
setting. The secondary infall solution for the density profile 
can also be connected to the nonlinear form of $\xi$ (Padmanabhan et al. 
1995; Padmanabhan 1996; Sheth \& Jain 1996). Here we
outline this connection,  and comment on its implications for the 
validity of stable clustering. This section is a bit of a detour from
the main body of the paper, and some readers may wish to skip to
sub-section 3.2 which provides a summary of the section. 

Consider a spherically symmetric distribution of 
collisionless particles with an initial density profile
\begin{equation}
{\delta\rho\over \rho} (x, t_i) \propto x^{-\kappa} \, .
\label{drhoinit}\end{equation}
Note that $\kappa$ here is $3 \epsilon$ in the notation of
Fillmore and Goldreich (1984) as they used the 
mass as the variable on the RHS. 
Let these particles be assigned velocities such that they
are in Hubble flow at some initial time $t_i$. The trajectories
of these particles will follow radial orbits, and since 
they are in overdense regions, they will eventually stop
expanding and collapse towards the center of mass. 
The point at which particles with an initial radius $x_i$
first stop expanding is known as turnaround, and the 
turnaround radius denoted $x_{ta}(x_i)$ depends on $x_i$. 
After turnaround, since the particles are taken to be collisionless, 
they execute oscillations about the center of mass. 

Fillmore \& Goldreich obtained self-similar solutions for the final 
distribution of these particles by using the adiabatic invariance
of the mass $M(x,t)$ within a given radius $x$. Consider the orbit
of a particle with maximum radius much smaller than the current
turnaround radius. By assuming $M(x,t)$
to vary sufficiently slowly that it could be treated as a constant
over the timescale of oscillation of this particle, 
they solved for the variation in time of the maximum
radius of the particle in terms of its turnaround radius. This in turn 
allowed them to obtain the asymptotic (late time) density profile 
in terms of the initial density profile. The result is:
\begin{equation}
{\delta\rho\over\rho}(x,t_{asym})\propto x^{\frac{-3 \kappa} 
{(1+\kappa)}} \qquad ; \qquad \kappa \geq 2 \, .
\label{fillmore}\end{equation}

The shape of the final density profile in the secondary infall model
is qualitatively different if $\kappa<2$, i.e. if the initial density 
profile is sufficiently shallow. The dynamical origin of this
difference is that for $\kappa<2$, at late times the mass within 
a radius $x$ is dominated by the contribution from particles
with maximum radii $>x$. In contrast, for $\kappa>2$, the mass 
within $x$ is dominated by particles with maximum radii $<x$. 
The consequence of this difference is that for $\kappa < 2$,
particle orbits on small scales keep shrinking due to the increasing 
mass contributed by large scales. 
The late time density profile approaches a form that does not depend 
on the initial profile. The asymptotic density profile is 
given by
\begin{equation}
{\delta\rho\over\rho}(x,t_{asym})\propto x^{-2}\qquad ;
\qquad \kappa < 2 \, .
\label{fillmore2}\end{equation}

The above discussion relies on purely radial orbits. The sharp
transition at $\kappa=2$ does not occur if one allows the
particles to have some angular momentum. White and Zaritsky (1992)
have studied the density profiles arising from an initial distribution
of angular momentum in which the eccentricity is constant as a function
of radius. In such a case, the transition to a regime of $x^{-2}$
density profiles does not occur at all, and the form of 
equation (\ref{fillmore}) is always valid. More generally 
one might find that a finite angular momentum causes the density 
profiles to lie somewhere in between the shapes of equation
(\ref{fillmore}) and (\ref{fillmore2}) for $\kappa<2$, and to
maintain the shape of equation (\ref{fillmore}) for $\kappa\geq 2$. 

\subsection{From secondary infall to stable clustering}

We are now in a position to apply the secondary infall model to 
the autocorrelation function $\xi$ in an Einstein-de Sitter
cosmology. The hope is that what works for an isolated halo also 
applies statistically to the full matter distribution with an 
appropriate matching of initial profiles. 
Consider the ingredients that led to the stable clustering
profile of equation (\ref{stable3}). 
(i) The pair conservation equation, which is 
simply a kinematical expression of the
conservation of particle number. (ii) Self-similarity, namely
that characteristic length scales grow as a power law of time:
$x\propto a(t)^{2/3+n)}$. (iii) Expanding coordinates, implicit
in the use of the comoving spatial coordiate $\vec x$. 

Each of these ingredients is present in the
secondary infall model described above. Thus stable 
clustering relies on such a limited aspect of
the dynamics of the full distribution, that the model of an
isolated spherically symmetric object can satisfy the conditions
under which it is applied. To connect the two models, we need
to relate the initial density profile
of the secondary infall model to the 
initial spectrum of the cosmological distribution. Unfortunately
there is no rigorous way to make this connection. We show below 
that the stable clustering slope for $\xi$ is recovered if the 
initial density profile is taken proportional to the rms smoothed density 
fluctuation of the cosmological distribution (Padmanabhan et al. 1995)
which is given by
\begin{equation}
{\frac{\delta\rho}{ \rho}} (x, t_i)\, \propto \, x^{-\kappa}\,
 \propto \, x^{-(3+n)/2} \, .
\label{drhoconn}\end{equation}

The Fillmore-Goldreich result for the final
profile given by equations (\ref{fillmore}) and
(\ref{fillmore2}) can now be directly adapted, 
with the substitution $\kappa=(3+n)/2$. Thus
the transition value of $\kappa=2$ corresponds to $n=1$, 
and the asymptotic shape of $\xi$ is
\begin{equation}
\xi(x,t_{asym})\, \propto\, x^{-{(9+3 n)\over (5+n)}}\, \qquad ;
\qquad  n\geq 1 \, ,
\label{drhoconn2}\end{equation}
corresponding to equation (\ref{fillmore}), and,
\begin{equation}
\xi(x,t_{asym})\, \propto \, x^{-2}\qquad ; \qquad n<1\, , 
\label{drhoconn3}\end{equation}
corresponding to equation (\ref{fillmore2}). 

Equation (\ref{drhoconn2}) shows one way 
to connect the stable clustering profile to
the nonlinear density profile of an isolated spherical halo. 
Since the choice of the initial profile was made in hindsight, 
in order to recover the desired shape of $\xi$, 
this is not intended to be  a ``derivation'' of 
stable clustering. What it does in fact demonstrate is that
there could be a regime in which stable clustering is {\it invalid}. 
With the particular initial profile chosen here, stable clustering
holds only if the initial spectral index $n>1$ -- a range of very little 
interest for realistic spectra like the CDM spectrum, since for
all scales of interest $n$ ranges between $1$ and $-3$ (or
lower for spectra with some hot-dark matter). Taken at face value
this result suggests that for spectra of interest 
stable clustering is invalid, and that the nonlinear $\xi$ should 
take the universal shape $\xi \propto x^{-2}$, independent of the
initial spectrum. 

However, as discussed above, the
presence of nonradial orbits lowers the transition value of $\kappa$
and of $n$, and possibly completely eliminates the 
regime of an $x^{-2}$ profile. The key parameter therefore becomes
the degree of eccentricity of the orbits -- if
the eccentricity stays roughly constant, or increases with the 
size of the orbit, then infalling particles will not contribute
to the mass inside a given $x$ and orbits
will not asymptotically shrink. In such a case, the profile will
retain the stable clustering prediction of equation
(\ref{drhoconn2}).  

One can also consider a more detailed model in which 
pairs with small separation $x$ are taken to belong to stable
halos, so that $\xi(x)$ arises from the convolution of the density 
profiles of such halos (Sheth \& Jain 1996). The stable clustering
shape of $\xi$ can be obtained from the secondary infall result
with an appropriate initial halo profile, 
but again there is no rigorous justification for this initial profile.

\subsection{Summary}

The lesson from the secondary infall models for $\xi$ is that they
show how the stability assumption can be invalidated in cases where
the initial profile is
sufficiently shallow. The key dynamical factor is the distribution
of angular momentum. Particles with large
orbital radii need  to have sufficient angular momentum, else their
contribution to the mass inside arbitrarily small radii causes
small scale orbits to asymptotically shrink.  This corresponds to a
net infall pair velocity, and could lead to a universal $x^{-2}$ shape
for $\xi$, independent of the initial $n$ for spectra of interest.  

In the formation of halos a more important process appears to
be the hierarchical merging of smaller halos. The influence of
merging on small scale dynamics can depend critically on the size 
of halos which contribute most of the pairs at small separations. 
If most of the pairs with separation $x$ belong to the cores of halos of 
radii much larger than $x$, then ongoing mergers of roughly equal mass
halos might cause the cores
to expand and lead to a net outflow pair velocity. On the other
hand if members of pairs come from halos with radii of order, or smaller
than, $x$, then the merging of such halos could lead to net infall
pair velocities. These considerations highlight some mechanisms
which could violate the stability assumption, and cause $\xi$ to
become either steeper or shallower than the stable clustering shape. 
But the limitations of analytical modelling leave N-body simulations
as the only realistic means of testing for stable clustering. 



\section{N-body simulations}

\begin{table*}
\caption{Parameters of the N-body simulations}
\medskip
\begin{tabular}{|l|c|c|c|c|}                  \hline

Simulation &   {\bf $N_{part}$ } & 
Softening $\epsilon/L$ & ${\rm PM}$ Mesh & $x_{min}/\epsilon$ \\  \hline 

$n=0$    & $100^3$ & $1/2500$ & $256^3$    & $2-4$\\ 

$n=-1$   & $100^3$ & $1/2500$ & $256^3$    & $2-4$ \\ 

$n=-1.5$ & $100^3$ & $1/2500$ & $256^3$    & $2.5-5$\\ 

$n=-2$   & $128^3$ & $1/2560$ & $256^3-432^3$ & $3-5$\\ 

CDM1     & $144^3$ & $65 {\rm Kpc}/100 {\rm Mpc}$ & $288^3-420^3$ & $2-3$\\ 

CDM2     & $100^3$ & $20 {\rm Kpc}/50 {\rm Mpc}$ &   $256^3$     & $3-4$\\ 
\hline
\end{tabular}
\label{Table}
\end{table*}

This section provides some details of the N-body simulations used
to obtain the results presented in the next section. 
Six different simulations are analyzed in this paper.
Two are for the standard CDM spectrum. The other four
are for power-law spectra with $n=0,$ $-1$, $-1.5$ and $-2$ with
$\Omega=1$.
All the simulations were performed using high resolution 
${\rm P^3M}$ codes. 
Table 1 shows some important parameters for the different simulations:
the total number of particles $N_{part}$, the Plummer force softening
parameter $\epsilon$, the size of the ${\rm PM}$ mesh, and the
smallest comoving scale $x_{min}$ on which $\xi$
can be reliably measured. 

For models with 
$-1.5\leq n\leq 0$, the simulations used were performed by S. White, 
with the same code as in EFWD. The total number of particles is
$100^3$ and the force resolution equivalent to a Plummer softening
parameter $\epsilon= L/2500$, where $L$ is the size
of the computational box.
The $n=-2$ run was performed by E. Bertschinger using 
an adaptive ${\rm P^3M}$ code. It followed $128^3$
particles with $\epsilon=L/2560$. 
The CDM simulation denoted CDM1 in Table 1, is the same as in
Gelb \& Bertschinger (1994).
It is a ${\rm P^3M}$ simulation with $144^3$ particles,  
$\Omega=1$, $H_0 =50 \, {\rm km/s/Mpc}$,  
$L=100\, {\rm Mpc}$ and $\epsilon=65\, {\rm kpc}$. 
It is normalized so that $\sigma_8$ (the linear rms mass fluctuation 
in a sphere of radius $16 {\rm Mpc}$) is unity when the expansion
factor $a=1$. 
For CDM at early times, $a<0.5$, we have used the results 
of a smaller box simulation performed by S. White. 
This simulation, denoted CDM2, has $100^3$ particles in
a $50\, {\rm Mpc}$ box, with a force resolution equivalent to
$\epsilon=20\, {\rm kpc}$. 

\subsection{Numerical Resolution Effects}

In using N-body simulations to study gravitational clustering, 
there are several numerical artifacts that need to be carefully
eliminated. In this sub-section, we focus on those effects that
are most relevant in studying the small-scale dynamics of interest
for this work. There are two critical factors: (i) The suppression
of $\xi$ on small scales due to force softening and 2-body relaxation,
and, (ii) Effects due to the initial conditions and the finite volume
of the simulation box which limit the range of time outputs and 
length scales over which clustering follows the true dynamics.
We provide estimates of the relevant numerical effects, and thus 
arrive at criteria for the range of scales on which the statistical 
measures used are reliable. For the scale free spectra these criteria 
are verified by testing for self-similar scaling. 

A comprehensive study and tests of numerical effects
can be found in CBH, Baugh et al. (1995) and Gelb (1992). These 
authors have tested the effects of finite resolution,
discreteness, and statistical fluctuations in the initial conditions
on different statistical measures. Their results on small
scales and on 2-point statistics are relevant here, as the simulations
used by them are of comparable resolution to those used in this study.
Tormen et al. (1996) have made detailed tests of numerical 
resolution in the context of measuring density profiles of halos --
these are also relevant as $\xi$ is closely related to the halo
profiles. 

\medskip
\noindent {\bf Initial conditions.} 
At early times the particle distribution is affected by the
cubical grid from which the particles are perturbed, and by
transients due to the Zel'dovich approximation used to set up the
initial perturbations. The effect of the grid can be minimized by
starting with ``glass'' initial conditions, as has been
done in the simulations for $-1.5\leq n \leq 0$ (see White 1994
for a discussion of ``glass'' initial conditions). 
In general, to ensure that there are no artifacts due to the initial 
conditions, the initial amplitude of the perturbations should be
chosen small. A test for how small is sufficient is provided by checking for
self-similarity, since the evolution of scale free initial spectra
is expected to be self-similar once it follows the true dynamics. 
The initial amplitude used in the scale free simulations with
$-1.5\leq n\leq 0$ is such that the power on the Nyquist
frequency of the particle grid equals the white noise level
(Efstathiou et al. 1988). For $n=-2$, as noted by Lacey \& Cole (1994), 
departures from self-similarity persist for rather late times. 
Hence the initial amplitude needs to be lower for $n=-2$; it
was chosen such that the dimensionless power on
the Nyquist frequency of the particle grid is $4 \pi k^3
P(k)=0.03$. 

\medskip
\noindent {\bf Force softening.} To suppress the effects of 
two-body relaxation, force softening is implemented in
computing short range forces in the particle-particle part of 
the simulations by using the equivalent of a Plummer force law: 
$F(r)=Gm^2r/(r^2+\epsilon^2)^{3/2}$. The parameter $\epsilon$ is given
in Table 1 for the different simulations used. 
The results shown in Figure 1 for $\xi(a,x)$ extend to 
$2 \epsilon$ on the small scale
end. As compared to the exact Newtonian force $G m^2/r^2$, 
the fractional error in $F(r)$ at $r=2 \epsilon$ 
is $28 \%$. The results in Figure 1 show that for the first 
one or two points plotted, in the range $2 \epsilon<x<3-4\epsilon$,
$\xi$ is underestimated, especially at earlier times. As discussed below, 
the cause for the artificial suppression of $\xi$ at earlier times 
is usually due to 2-body relaxation and not force softening.

\medskip
\noindent {\bf Discreteness effects: 2-body relaxation.} 
The effect of force softening is easy to quantify, as described
above, and since $\epsilon$ is constant in comoving coordinates, 
it is expected to affect the same range of scales at all times. 
But the distribution at separations $>2 \epsilon$ 
may still be inaccurate if randomly chosen particles do not have
enough neighbours within such a separation, due to the effects
of 2-body relaxation. A
simple estimate can be made because $1+\xi(x)$, when 
integrated over a sphere of radius $x$, is the mean
number of neighbours 
within separation $x$ from a randomly chosen particle. 
For $\xi$ large enough this is essentially the total number of
neighbours. Therefore with $x=m \epsilon\simeq m
L/2000$, where $m$ is of order $2-5$, and
$N_{part}^3/L^3\simeq 10^6/L^3$ being the mean number density, 
we can estimate the mean number of neighbours to be: 
\begin{equation}
N_{nbrs} = \frac{4 \pi}{3-\gamma} \left(\frac{N_{total}}{10^6}\right) 
\left(\frac{\xi}{10^3}\right) \left(\frac{m}{2}\right)^3 \, .
\label{nbors}\end{equation}
In obtaining the above equation, we have assumed that $\xi\propto
x^{-\gamma}$ down to $x=0$, with $\gamma=1-1.8$ for the spectra of
interest. In the simulations however, $\xi$ is
suppressed on very small scales. The extreme case is given by
setting $\gamma=0$ which decreases $N_{nbrs}$ by a factor of
$\simeq 2$. 

We use the empirical criterion that $N_{nbrs} > 10$ is required to measure 
$\xi$ accurately. The resulting minimum scale $x_{min}$ is given 
in Table 1  and is $> 2 \epsilon$ at the 
two earlier time outputs shown in Figure 1 (in stars and circles). 
It is typically smaller than $2 \epsilon$ for the last output, 
in which case we set $x_{min}=2 \epsilon$ due to the effect of
force softening. Taking the minimum reliable scale to be the 
larger of the values that satisfy the criteria of accurate forces
and sufficient particle number
successfully identifies the small scale regime in which 
the measured $\xi$ departs from self-similarity. 
For CDM we have checked that the measured $\xi$ in
the two simulations, which have different values of $N_{nbrs}$
at a given scale, are in very good agreement
on the scales used in our analysis (see Figure 4).
Our results for $x_{min}$ are similar to those of CBH after adjusting 
for the different particle number and force softening.

\medskip
\noindent {\bf Range of $x_{nl}(a)$.} At late times the nonlinear
scale $x_{nl}(a)$, defined by equation (\ref{xnl}),
approaches the size of the box. This
sets a maximum value of $a$ beyond which the clustering
is artificially suppressed due to
the absence of modes with wavelength larger than the box-size.
This problem is particularly severe for spectra with $n\leq -2$ 
because they have more power on large scales
relative to spectra with higher $n$, which in turns leads to a
stronger nonlinear coupling with small scale modes. 
It is the principal reason for the difficulty in obtaining
self-similar scaling for the $n=-2$ spectrum. 
For spectra with $n>-2$ we find that $x_{nl}(a)>L/10$ is sufficient
to obtain accurate self-similar scaling on the scales of interest. 
This is the criterion used to set the maximum time output for
our analysis.

At very early times there are transient effects from the initial
conditions, and moreover clustering on small scales has not fully
developed. We allow for an expansion factor of at least $4$, 
and in the case of $n=-2$ of $12$, between the initial time 
and the first output used for our analysis. At this time the 
nonlinear scale is in the range $x_{nl} \simeq L/50-L/100$.

\bigskip
\noindent {\bf Other effects.}\newline
{\it ${\rm PM}$ mesh:} 
The mesh used to compute the long range forces in the ${\rm PM}$
(particle-mesh) part of the force computation also introduces
an artificial scale. But if the particle-particle forces are computed for a
conservatively chosen set of particles at each timestep, as is done
for the simulations used here, then the effects of the mesh are 
virtually absent (Bertschinger 1991). \newline
{\it Poisson fluctuations:} 
To compute $\xi$ at a given radius, sums over particles within 
logarithmically spaced radial bins are made. 
Poisson fluctuations within each radial bin produces errors in 
the measured $\xi$. This effect is also negligible as the number
of pairs measured is large enough that the resulting fluctuations 
in $\xi$ are smaller than $1\%$, and would appear smaller than
the symbols shown in Figures 1 and 2. \newline
{\it Finite volume effects:} At low wavenumbers
the finite number of modes present in a given bin in $k$ causes 
statistical fluctuations in the initial power spectrum. As nonlinear 
evolution proceeds, these low$-k$ modes couple to the modes at
higher $k$ and can in principle lead to fluctuations in the 
distribution at small scales as well. Also, modes with wavelength
larger than the box-size are absent. In practice for the 
range of $x_{nl}(a)$ used, there is negligible effect on scales
smaller than $L/10$, which is the range of scales of interest here. 
The effect is not negligible on the higher moments of the density, as
emphasized by CBH and Baugh et al. But for the evolution of $\xi$ 
the fluctuations between different realizations are about $2-3 \%$
on the  small scales. We have checked this by comparing $\xi$
for simulations started from three different realizations of the 
$n=-1$ spectrum.

\begin{figure*}
\centering
\epsfxsize=\hsize\epsffile{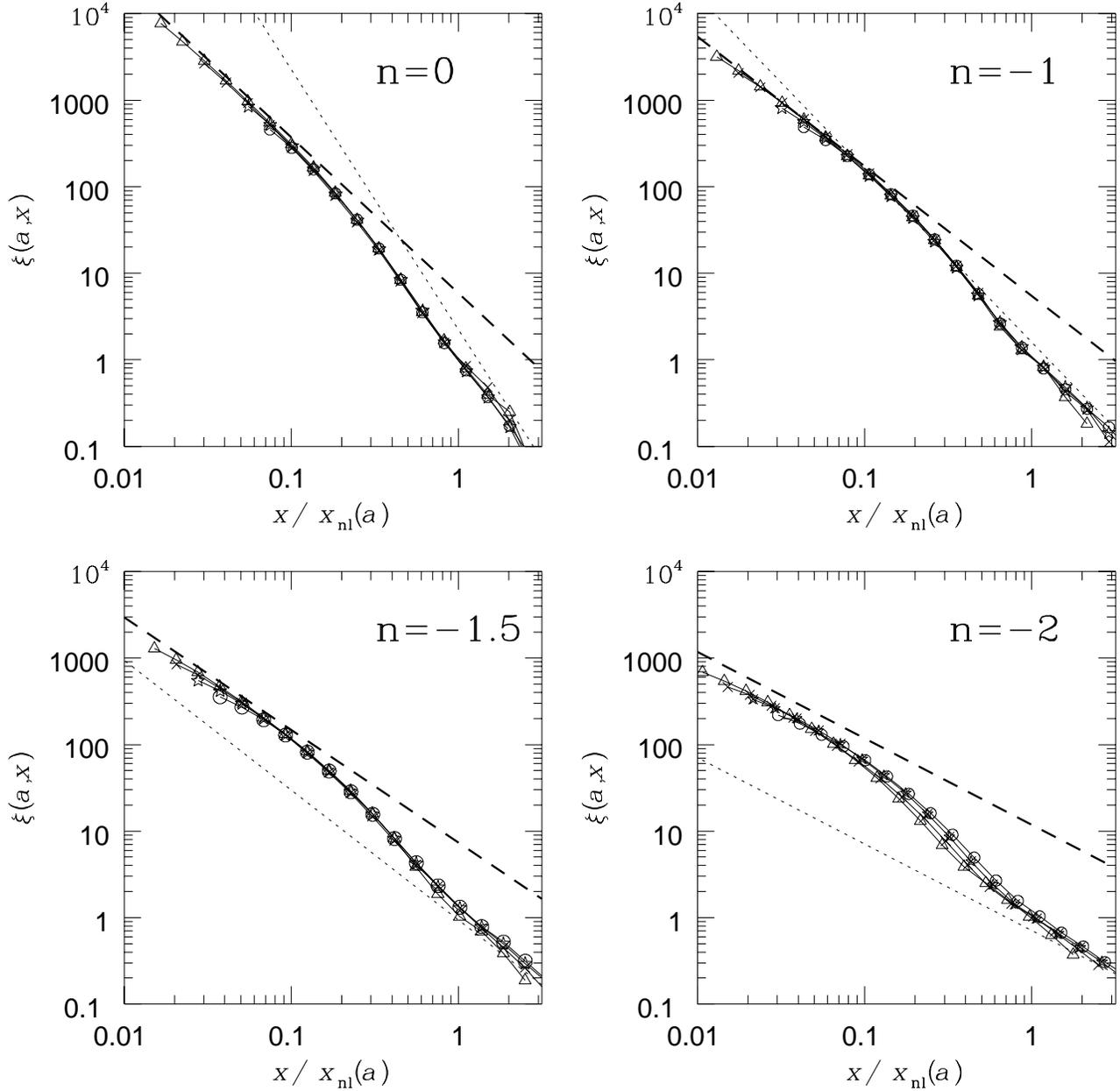}
\caption{Self-similarly scaled correlation functions. Each panel shows
$\xi(a,x)$ computed at four time outputs from simulations of scale
free spectra with $n=0,-1,-1.5,-2$ as indicated in the panels. It is
plotted against $x/x_{nl}(a)$, where $\xi(a,x_{nl})=1$ defines the
nonlinear scale $x_{nl}(a)$ at each $a(t)$. Between successive time outputs 
$x_{nl}(a)$ grows by a factor of $\simeq 1.5$. The circles, stars, 
crosses and triangles show the outputs at successively later
times, and therefore at later stages of nonlinearity. The dotted line
is the linear $\xi$. The
dashed line shows the stable clustering slope for $\xi\, $;
its amplitude is given by the model of Sheth \& Jain (equations
19 and 20). 
}
\label{corr1}
\end{figure*}

\section{Results}

Self-similarity provides a powerful simplification in 
studying the evolution of scale-free spectra in an Einstein-de 
Sitter universe. Using the self-similar solution for $\xi(a,x)$
from equation (\ref{stable3}), we define
a characteristic nonlinear scale $x_{nl}(a)$ as
\begin{equation}
\xi(a,x_{nl})\equiv 1 \qquad ; \qquad x_{nl}(a) \propto a^{2/(3+n)} .
\label{xnl}\end{equation}
In the results presented in this section, the comoving distance
$x$ will be scaled by $x_{nl}(a)$ at each time. Expressed as a 
function of $x/x_{nl}(a)$ the correlation function $\xi$, or any
dimensionless statistic, is independent of time  if the evolution
is self-similar. This allows the measured correlation function at 
different times to be combined so that the effective dynamic range of 
the simulation is extended. 
See Jain \& Bertschinger (1996) for an analytical analysis, 
and CBH and Padmanabhan et al. (1995) for numerical tests of 
self-similar  evolution.

Figure 1 shows $\xi(x/x_{nl}(a))$ for four different power law
spectra. In each panel four output times are used to compute $\xi$.
The figure shows that  self-similar scaling is convincingly 
verified for all the spectra, with the possible exception of the
$n=-2$ case which shows some scatter. Small departures from
self-similarity can be seen at the smallest and largest scales. 
On small scales the departures arise because $\xi$ is underestimated
due to the effects of force softening and finite particle
number. The effect is very small for all points except for the 
the very last, which corresponds to a separation $\simeq 2 \epsilon$. 
On large scales the correlation function is affected,
especially at late times, by the absence of waves that are larger 
than the size of the box. This causes $\xi$ at late times to fall
below the self-similar curve. 

As pointed out by CBH, a simple way to estimate the self-similar
$\xi$ for a given spectrum, is to take the largest of the different
values measured at different times for each value of $x/x_{nl}(a)$. 
This should give a reasonable result for the cases where there is
some scatter, as in $n=-2$, or the small scale end of all the
spectra. The rationale is that numerical limitations cause a
suppression of $\xi$, both at small and large scales. However, we
shall not need to do this because for the scales of interest, the
measured $\xi$ are sufficiently self-similar without any corrections. 

With the self-similarly scaled correlation functions for different
spectra, the prediction of stable clustering can be tested. The
dashed lines in each of the panels of Figure 1 show the slope
predicted by stable clustering. It is evident that on the smallest
scales, for about a factor of $5$ in $x/x_{nl}(a)$, the measured slope 
of $\xi$ agrees with the stable
clustering slope. As a first approximation this occurs for $\xi \gg
100$ for all spectra. A more detailed examination of the slope of $\xi$ in 
Figure 1 and of the mean pair velocity in Figure 3 
shows that the onset of stable
clustering occurs at higher $\xi$ for spectra with larger $n$; thus
for $n=-1.5$ stable clustering is valid for $\xi\gsim 200$, whereas
for $n=0$, it is valid for $\xi\gsim 500$. 

Interestingly, the amplitudes of $\xi$ which demarcate the onset of
stable clustering correspond to approximately the same value of $x/x_{nl}$
independent of the initial spectrum: $x/x_{nl}(a)\lsim 0.07$. 
This provides a convenient demarcation of the stable
clustering regime as it is independent of the spectrum or of any
feature of self-similarity, and can therefore be 
applied to realistic spectra like the CDM spectrum. The variation
with $n$ in the value of $\xi$ corresponding to $x/x_{nl}(a)\lsim
0.07$ arises because spectra with larger $n$ have steeper $\xi$
in the regime of $0.07<x/x_{nl}<1$. Since $\xi(x_{nl})=1$ by
definition, $\xi(0.07 x_{nl})$ increases with $n$. 

The amplitude of the stable clustering $\xi$, i.e. the normalization
of the dashed line in Figure 1, has been fixed by the analytical
model of Sheth \& Jain (1996). Their calculation assumed that on 
small scales $\xi$ is determined by the density profiles
of collapsed halos. They fixed the shape of the profile by assuming
the validity of stable clustering, and used the
Press-Schechter form for the number density of halos to then predict the 
amplitude of $\xi$. The final result, equation (16) of Sheth \& Jain,
can be re-written as
\begin{equation}
\xi(x/x_{nl})\, =\, C_n\, \left(\frac{x}{x_{nl}}\right)^{-\gamma}\, ,
\label{sheth}\end{equation}
where $\gamma=3(3+n)/(5+n)$ is the stable clustering slope, and 
$C_n$ is an $n-$dependent constant. For the spectra used in this
paper it takes the values, 
\begin{eqnarray}
C_n&=&\{5.93,5.40,7.42,11.96\} \qquad {\rm for} \nonumber \\
n&=&\{0,-1,-1.5,-2\} ,
\label{cn}\end{eqnarray}
respectively. 

The results show that again, except possibly 
for $n=-2$, the measured $\xi$ is in remarkably good agreement with
the predicted value. For the range of small scales with $\xi>200$, 
the measured $\xi$ for $-1.5\leq n \leq 0$ is within $\simeq 10 \%$ of the 
prediction. For $n=-2$ the predicted value is larger by $\simeq 30
\%$ for the maximum measured $\xi$. For this case, the large scatter
between the different time outputs 
suggests that even the maximum measured $\xi$ at a given $x/x_{nl}$ 
is an underestimate of the true value; the dashed line might
therefore be accurate at the same level as the other cases. 
The verification of the amplitude of $\xi$ predicted by 
equation (\ref{sheth}) implies that
both stable clustering and the Press-Schechter model provide accurate
descriptions of nonlinear clustering for a wide range of spectra. 
It also suggests that it is reasonable to assume that $\xi$ is
determined by the structure of
dark halos on sufficiently small scales. The trend that the onset of stable 
clustering occurs for larger values of $\xi$ as $n$ increases is also
in agreement with the estimate of Sheth \& Jain, which assumed 
that only the half-mass radii of virialized halos have stabilized. 
Our results can be used to estimate the free parameter in the 
spherical model for $\xi$ used by Padmanabhan et al. (1995). 
The choice $m=3$ in their equation (14) gives closest agreement
with our results. 
 
\subsection{Mean pair velocity}

\begin{figure}
\epsfxsize=\hsize\epsffile{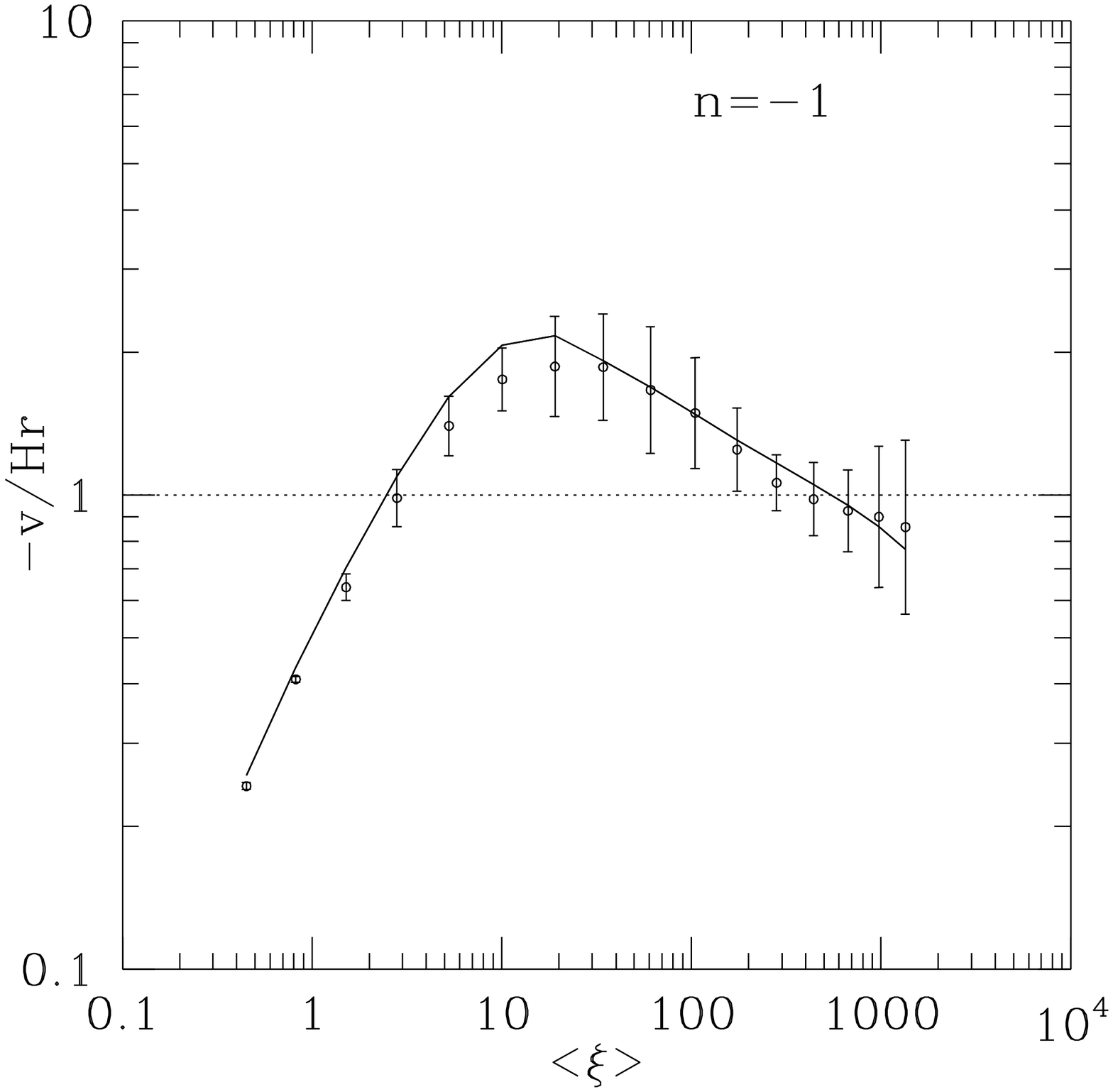}
\caption{Mean pair velocity for $n=-1$. Simulations of three different
realizations of the $n=-1$ spectrum are used to directly measure
$-v/Hr$. The average value is shown by the circular data points, with
the error bars showing the $1$-$\sigma$ deviations of the
distribution. The solid curve 
shows $-v/Hr$ computed using the pair conservation equation as in
Figure 3, for the same output time as the circular points. 
}
\label{pair2}
\end{figure}

The stable clustering hypothesis can also be tested by directly
computing the mean relative velocity of pairs of particle. In physical
coordinates this velocity should approach zero as the pair separation
is made vanishingly small. In comoving coordinates therefore, the
mean pair velocity $v(a,x)=a\dot x$ must exactly cancel the Hubble 
velocity $Hr=\dot a x$, so that the ratio $-a\dot x/Hr=1$. 

However $v(a,x)$ is much more noisy than $\xi(a,x)$, because of
the high dispersion in the pair velocity. This is illustrated in
Figure 2 where the measured $-v/Hr$ is plotted against $\bxi$
for $n=-1$.  It is only on averaging the data from three realizations
that a meaningful result is obtained. At least five
realizations for each spectrum would be needed to directly measure 
$-v/Hr$ to better than $20\%$. In the absence of such extensive 
simulation data, we use an indirect method to measure $-v/Hr$. 

\begin{figure*}
\epsfxsize=\hsize\epsffile{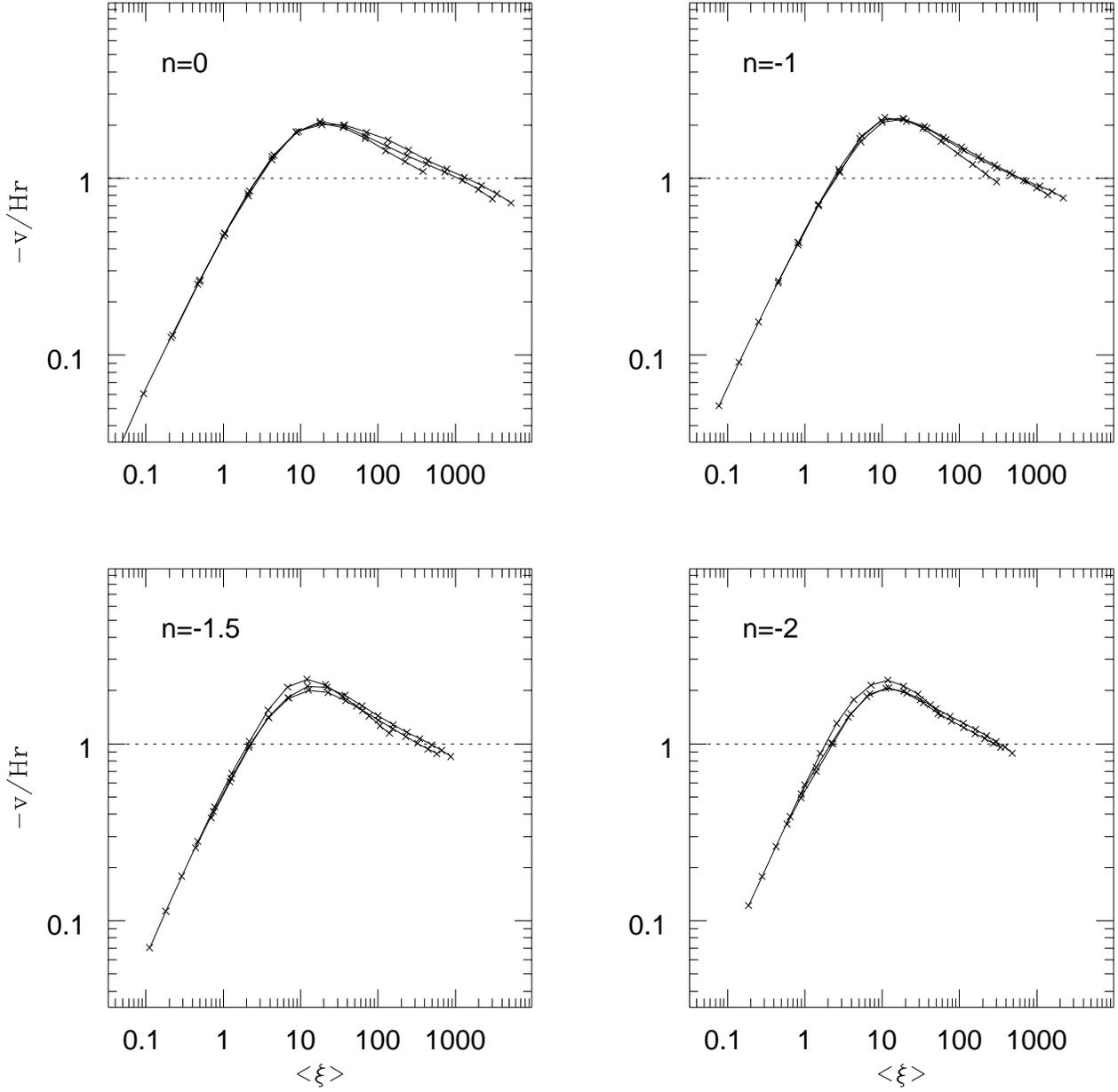}
\caption{The ratio of the mean pair (peculiar) velocity to the Hubble 
velocity, $-v/Hr$, as a function of the mean correlation function $\bar
\xi$. The pair conservation equation is used to solve for $-v/Hr$ 
using the evolution of $\bxi(a,x)$, as discussed in the text.
Results are shown for three output times,
corresponding approximately to the outputs shown in Figure 1, for the same 
set of scale free spectra. The near coincidence of the
curves at different times verifies self-similarity, while the approach
to the dotted line $-v/Hr=1$ is consistent with stable clustering.  
}
\label{pair1}
\end{figure*}

As described in Section 2, the evolution of $\xi$ is connected
with $v$ via the pair conservation equation. Therefore to the
extent that the simulation is accurate, $v$
can be measured using the evolution of $\xi$, and should give the
same result as would be obtained from a simulation with enough particles
to reduce the noise in the directly measured $v$. We use
the form of the pair conservation equation given in equation 
(\ref{stable4}) to solve for $v$. We evaluate the time derivative of
$\bxi$ by taking logarithmic derivatives. Since the evolution
of $\xi$ and $\bar \xi$ is close to a power law in $a$, taking
logarithmic derivatives reduces the error introduced by the finite
time interval between successive outputs. 
To estimate $-v/Hr$ for a given $x$ at
output time $a_i$ we use the finite difference equation
\begin{equation}
\frac{-v} {H r}(a_i)=
{\bxi(a_i)\over
3[1+\xi(a_i)]}\left[\frac{\log\bxi(a_{i+1})-\log\bxi(a_{i-1})} 
{\log a_{i+1}-\log a_{i-1}} \right] \, .
\label{pairnbody}\end{equation}
While the mean pair velocity contains the same information
as the evolution of $\xi$, it tests for
stable clustering more directly for generic spectra like CDM
which do not evolve self-similarly. 

The results for the inferred $-v/Hr$ are shown in Figure 3
for three time outputs of the scale free spectra with
$n=0,-1,-1.5,-2$. Self-similarity requires
that the data at different time outputs coincide, since both
$-v/Hr$ and $\bxi$ are dimensionless functions of $x/x_{nl}(a)$.
On scales with $\bxi<1$ the linear 
theory relation $-v/Hr \propto \xi$ is verified. For
larger values of $\bxi$, the ratio $-v/Hr$ turns over after
reaching a maximum value of $\simeq 2$ at $\bxi\simeq 10$. 
For $\bxi>100$, the ratio approaches unity, though it appears
to continue to decline with increasing $\bxi$ for a given time. 
For the range of $\xi$ for which the slope in Figure 1 agrees 
with the stable clustering value, the measured $-v/Hr=1$ to within
about $20\%$ for almost all the data. While this is a 
satisfactory consistency check, 
$-v/Hr$ does not appear to flatten into an asymptotic regime
up to the highest $\bxi$ measured. 

The results for $-v/Hr$ on small scales are less reliable
than the results for $\xi$ because the former depends directly 
on $\bxi(x)$, which is an integral of $\xi$ over scales smaller than $x$, 
and is therefore suppressed even for $x>x_{min}$. 
This artificial suppression largely accounts for the continued decline
in the measured curves below the line $-v/Hr=1$. We verified that if
$\xi(x<\epsilon)$ is fixed at the stable clustering slope, then
on scales in the range $2\epsilon <x<4 \epsilon$, shown by the last 
3 points at each time output, $-v/Hr$ would indeed be almost flat. 
Larger simulations, which accurately measure 
$-v/Hr$ on scales significantly smaller than in our
simulations, are needed  to test its approach
to the line $-v/Hr=1$. 

While the evidence for the existence of an asymptotic regime in 
Figure 1 for $\xi$ is still weak, 
the results for $\xi$ and the mean pair velocity 
are certainly consistent with stable clustering on the
smallest scales probed. These results are in agreement with those of CBH,
who used tree-code simulations in their analysis, with force
resolution comparable to our simulations and $< 1/4$ the number 
of particles. Padmanabhan et al. (1995) however concluded that 
for $\Omega=1$, stable clustering is violated as they obtained
$-v/Hr > 1$. A careful comparison of their results with ours
reveals that the results from their ${\rm P^3 M}$ simulations are consistent
with ours, and with stable clustering. But the results
for $n=-2$ and $n=-1$ from their ${\rm PM}$  simulations are inconsistent
with ours at the maximum $\xi$ measured in their
simulations (typically a factor of $2-3$ smaller than ours). 
The reasons for this discrepancy are not clear, and merit a detailed
comparison of clustering in high-density regions in ${\rm PM}$ 
and ${\rm P^3M}$ simulations. 

\subsection{Evolution of the CDM spectrum}

\begin{figure}
\epsfxsize=\hsize\epsffile{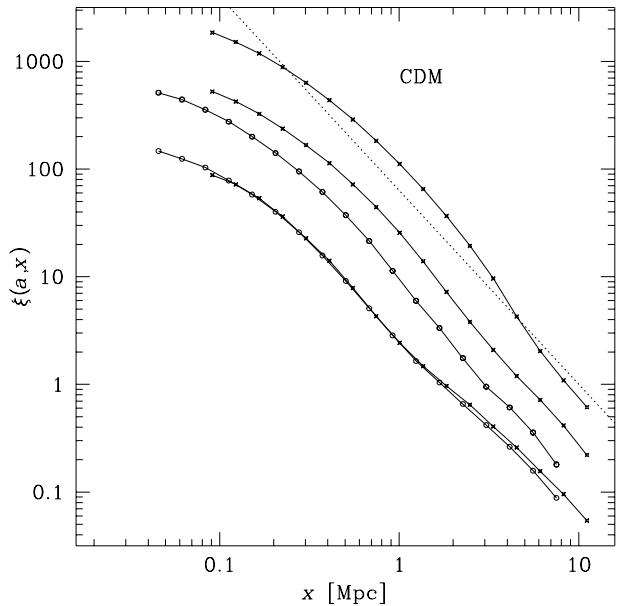}
\caption{Correlation function for the CDM spectrum. 
The solid curves show $\xi(a,x)$ computed from two simulations
of the standard Cold Dark Matter power spectrum 
at four output times, with $a=0.21,0.32,0.51,1.0$. The data
shown in circles are from a simulation with a box size $L=50\ {\rm
Mpc}$, and those in stars are from a bigger box with $L=100\ {\rm Mpc}$. 
To check for the accuracy of the measured $\xi$, the data at $a=0.2$
from both simulations are shown. The dotted line shows the observed
galaxy auto-correlation function with slope $= -1.8$. 
}
\label{corr2}
\end{figure}

\begin{figure}
\epsfxsize=11cm \epsfysize=13cm
\epsffile{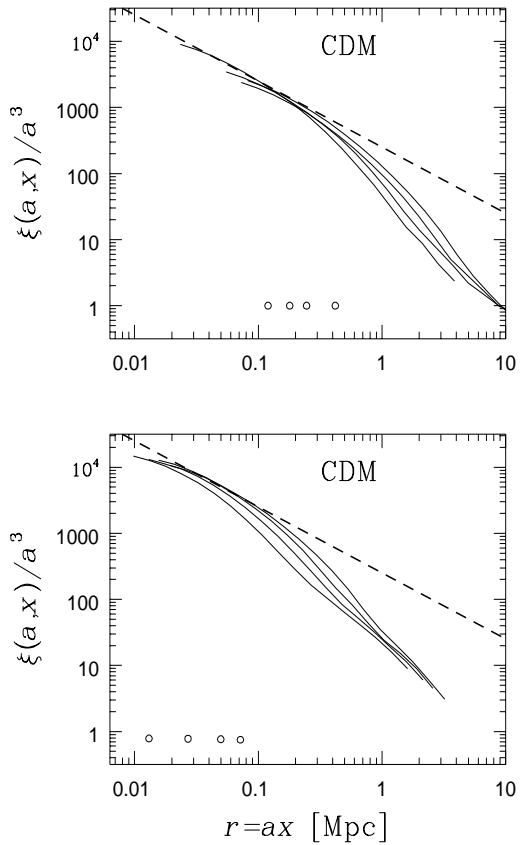}
\caption{Scaled correlation function for the CDM spectrum. 
The two panels show  $\xi(a,x)/a^3$ for the CDM spectrum
plotted against the physical length scale $r=ax$. The different curves
should coincide in the stable clustering regime. The upper panel
shows expansion factors $a=0.51,0.60,0.81,1.0$, and the lower panel
shows $a=0.21,0.28,0.34,0.43$. The data for $a>0.5$ and $a<0.5$ are
separated as the numerical resolution limit on small scales precludes
their comparison over a common range in $r$ in the stable clustering regime. 
The slope and amplitude of the dashed lines in both panels 
are the same as the stable clustering values for $n=-2$ shown in
Figure 1. The x-coordinate of the circles marked towards the bottom 
is the scale $0.07\, r_{nl}(a)$; for given $a$ stable clustering
is expected for $r< 0.07 \, r_{nl}(a)$ as discussed in the text. 
}
\label{corr3}
\end{figure}

The results for $\xi$ and $v$ for 
the standard CDM spectrum with $\sigma_8=1$ at $a=1$ 
are shown in Figures 4-8. Figure 4 shows $\xi$
measured at 4 different redshifts in the range $0.2<a<1$. We use
results from both simulations to maximize the range of 
scales on which $\xi$ can be accurately measured. At early times, 
for $a<0.5$, when the correlation length $x_{nl}<5 \ {\rm Mpc}$, we 
measure $\xi$ from the simulation with a smaller box-size, CDM2, 
as it has better resolution on small scales. 
For $a>0.5$ CDM1 is used as $x_{nl}$ becomes larger than $L/10$ for
CDM2. We are thus able to measure $\xi$ down to $x_{nl}/30$ for 
$a=0.2$, and $x_{nl}/100$ for $a>0.4$. Since the results of 
Figure 1 show that $x/x_{nl}<0.07$ is required for stable clustering, 
we have a sufficient range of $x$ to test the CDM model in the 
range $0.2<a<1$. 

Unlike the scale
free spectra, for CDM it is not possible to scale self-similarly
and compare $\xi$ at different times. However, if stable
clustering is valid, then expressed as a function of the separation
$r=ax$ in physical coordinates, $\xi(a,r)$ must grow in time as $a^3$
(see equation (\ref{stable2})) so that the mean number of neighbours
$\propto a^{-3} \xi(a,r)$ remains constant. We therefore test for the
constancy of $a^{-3} \xi(a, r)$ on sufficiently small scales 
in the two panels 
of Figure 5 for four values of the expansion factor in each panel. 
Since the onset of stable clustering should occur approximately
at a fixed value of $\xi$, the physical scale at which this occurs
increases with $a$. Figure 6 shows the 
evolution of the correlation length $r_{nl}(a)$ with $a$. The values of
$r=0.07\, r_{nl}(a)$ for the four values of $a$ in each 
panel of Figure 5 are shown by the circles at the bottom. 
Thus for the curve extending to the smallest $r$, the onset of
stable clustering is marked by the left-most circle.  
The values of $r=0.07 \, r_{nl}(a)$ are $20, 180, 700 \, {\rm kpc}$
for $a=0.2, 0.5, 1$ respectively. The rapid growth of $r_{nl}(a)$ 
makes it difficult to assess for stability for non-power law
spectra by comparing $\xi$ at different times. 
But within the region of overlap to the left of the 
circles, the different curves are very close to each other. 
The results are therefore consistent with stable clustering, as
verified also by the plot of $-v/Hr$ in Figure 8. 

Both the amplitude and slope of $\xi$ on small scales 
are very well approximated by the
dashed line in Figure 5, which is the same as for the $n=-2$ spectrum
in Figure 1. Thus its slope is $-1$ and its amplitude is set by equations
(\ref{sheth}) and (\ref{cn}) 
with $n=-2$. For the lower panel in Figure 5, with 
$a<0.5$, the measured slope of $\xi$ is shallower, as it is sensitive
to the part of the CDM spectrum with $n<-2$. However the $n=-2$ line
is still an adequate approximation. We conclude that for CDM-like spectra the 
slope and amplitude of $\xi$ on small scales is not very
sensitive to the precise shape of the spectrum, and is approximated
rather well by  the stable clustering result for a power law
spectrum with $n\simeq -2$. 

\begin{figure}
\epsfxsize=\hsize\epsffile{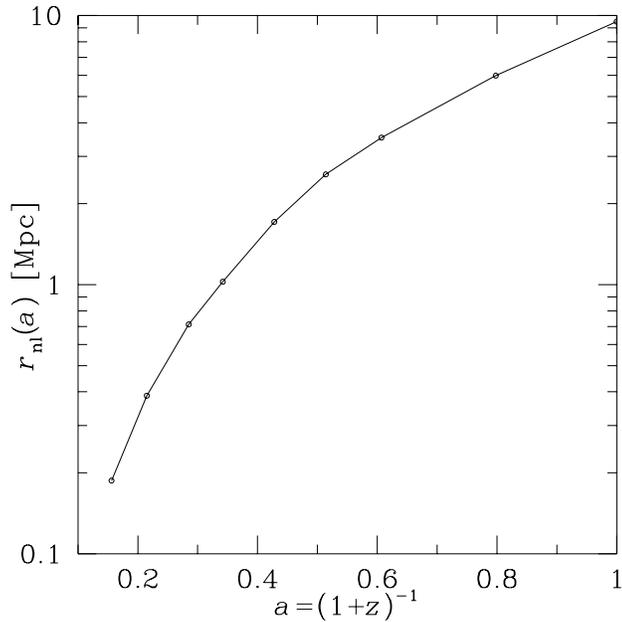}
\caption{Correlation length for the CDM spectrum. 
The measured correlation length in physical coordinates, defined by
$\xi(a,r_{nl})=1$, for the CDM spectrum is shown for $0.15<a<1$. 
The scale $0.07\, r_{nl}(a)$, shown in Figure 5 to mark the onset of stable 
clustering, can be read off from this plot for a given $a$. 
}
\label{corrlength}
\end{figure}

\begin{figure}
\epsfxsize=\hsize\epsffile{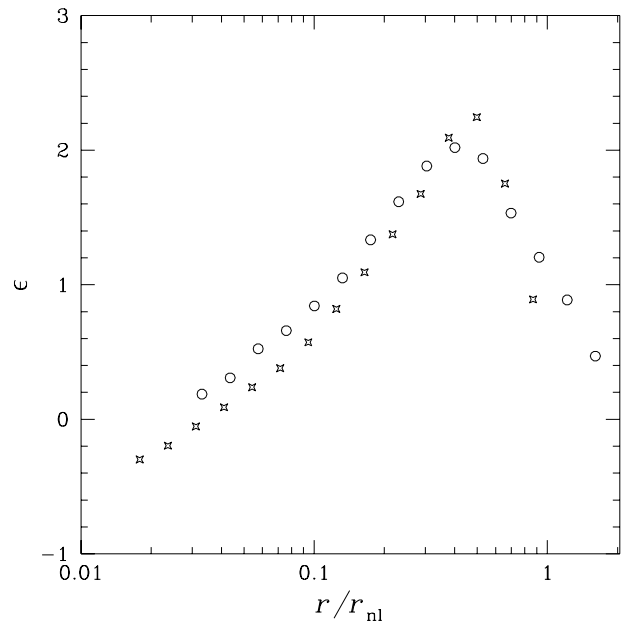}
\caption{Growth of $\xi(a,r)$ for CDM. 
For a power law parameterization of the form
$\xi(a,r)=a^{3+\epsilon}\, 
\tilde{\xi}(r)$, 
$\, \epsilon$ is shown as a function of $r/r_{nl}$. The circles and
stars are for $a=0.28$ and $a=0.51$ respectively. The variation
of $\epsilon$ with $r/r_{nl}$ shows that a simple power law form
for all $r$ is not accurate, and that in the regime $0.1<r/r_{nl}<1$, 
the growth rate is much faster than the commonly used values of 
$\epsilon=0$ or $0.8$. The value of $\epsilon$ can be obtained as a 
function of $r$ for any desired $a=1/(1+z)$ by using 
Figure 6 to get $r_{nl}(a)$.
}
\label{epsilonfig}
\end{figure}

\begin{figure}
\epsfxsize=\hsize\epsffile{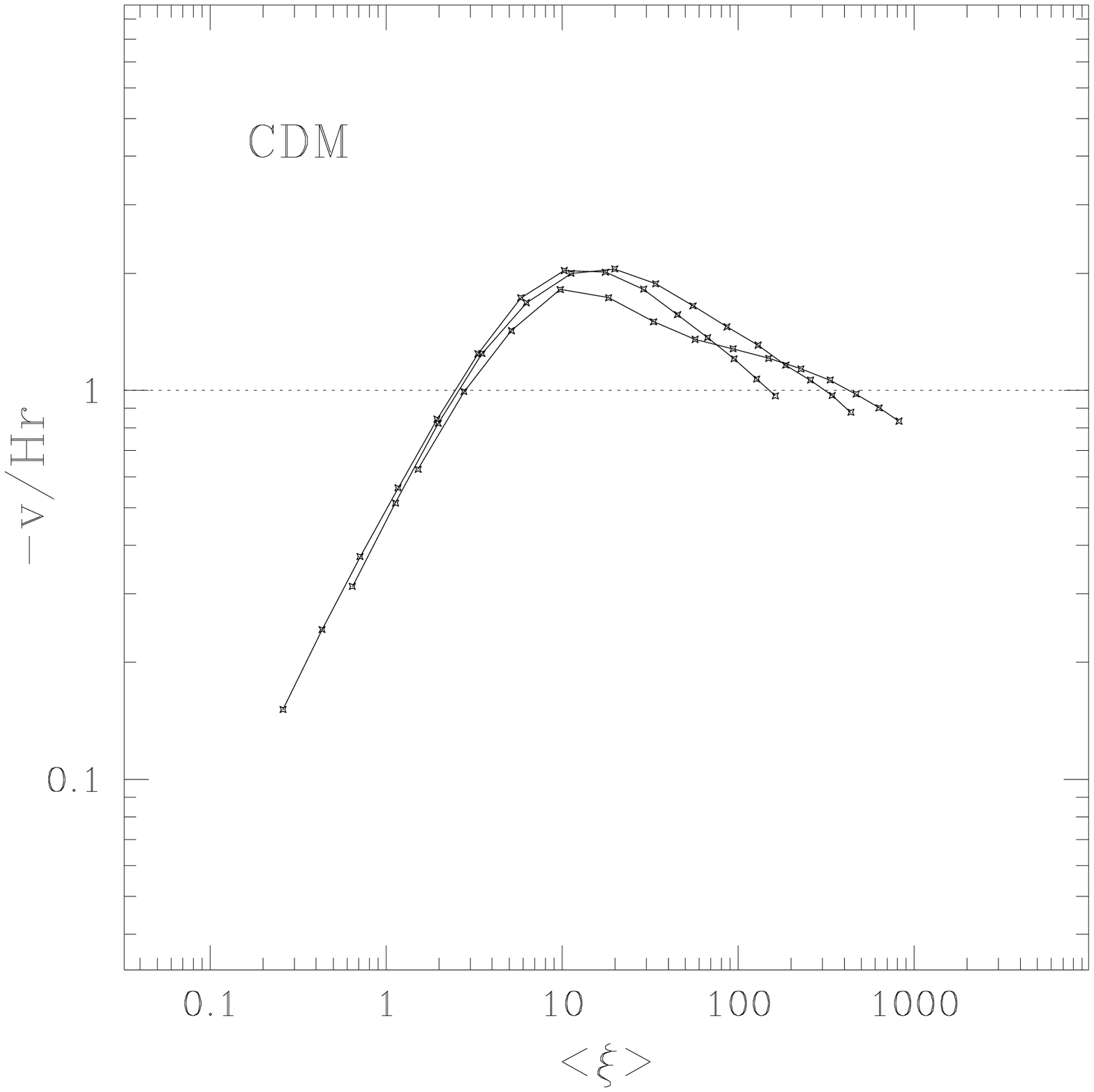}
\caption{Mean pair velocity for the CDM spectrum. 
The mean pair velocity is computed using the pair conservation 
equation as in Figure 3. The three curves are for $a=0.3, 0.6, 0.8$. 
Note that unlike Figure 3, for CDM-like spectra
the curves at different times are not expected to coincide. 
They do however approach the stable clustering value $-v/Hr=1$ 
for $\bxi>200$. 
}
\label{pairconfig}
\end{figure}

Finally we test a parameterization of the evolution of $\xi$ 
commonly used in studies of clustering of galaxies and Lyman-alpha
clouds at high redshift. The correlation function is taken to
have the form:
\begin{equation}
\xi(a,r)=a^{3+\epsilon} \left(\frac {r}
{r_0}\right)^{-\gamma_0}\, ,
\label{param}\end{equation}
where $\gamma_0=1.8$ is the observed slope of the galaxy correlation
function, and $r_0\simeq 5 h^{-1} {\rm Mpc}$ is the correlation 
length at $a=1$. The parameter $\epsilon=0$ in the stable clustering regime
and $0.8$ in the linear regime. Our results show that there are two 
main reasons why such a parameterization is inaccurate in the context
of any CDM-like model. First, the growth of $\xi(a,r)$ in the intermediate
regime is much faster that $a^3$, so the stable clustering and linear
regime do not give the upper and lower bound for the growth rate of
$\xi$. Second, the onset of the stable clustering regime, and of the
intermediate regime as well, occurs on a scale which grows rapidly
with $a$. Therefore to parameterize the shape of $\xi$ by one or two 
power laws which remain fixed in time is not accurate. 
For general use, following Hamilton et al. (1991) and
Nityananda \& Padmanabhan (1994),
universal fitting formulae have been proposed for the nonlinear $\xi$
which apply to any smooth initial spectrum 
(see Jain, Mo \& White 1995 and Peacock \& Dodds
1996). 

However, if for reasons of convenience, a parameterization
like that of equation (\ref{param}) is used, the results of
Figures 5  and 7 show that $\epsilon=0$ is a good approximation for 
$r/r_{nl}<0.07$, with $r_{nl}(a)$ shown in Figure 6. In the regime
$0.07<r/r_{nl}<1$ the growth rate varies rapidly, rising to $\epsilon\simeq
2$ before approaching the linear theory value $\epsilon=0.8$ at
$r> r_{nl}$. Thus in the intermediate regime between the stable
clustering and linear regimes, $\epsilon$ is typically larger than $1$,
and is significantly underestimated if it is taken to be $0$. 

Figures 6 and 7 can be used to estimate $\epsilon$ for a CDM-like
model in which the growth of $\xi$ is parameterized as: $\xi(a,r)=
a^{3+\epsilon}\, \tilde{\xi}(r)$, with $\tilde{\xi}$ being a function
only of $r$ in some narrow range of $a$.   
Figure 6 gives $r_{nl}(a)$ for any desired $a=1/(1+z)$, and
Figure 7 can then be used to read off $\epsilon$ at the range of
physical separations $r$  of interest. For CDM-like spectra and 
with $\Omega=1$, the 
results of Figure 7 are not sensitive to the shape of the
spectrum and to the choice of $a$ in the range $0.2\lsim a \lsim 1$. 
This discussion is of course
subject to the uncertainties due to the possible biasing of galaxies
relative to the dark matter. 

\section{Discussion}

The results presented in this paper can be summarized as
follows. 

\noindent $\bullet$ The shape and evolution of the correlation function 
$\xi(a,x)$ is consistent with the stable clustering prediction
of equation (\ref{stable3}). We have verified this result
with the power spectrum measured from the simulations
presented here as well. 

\noindent $\bullet$ Direct measurement of the mean pair velocity 
is not sufficiently accurate on small scales. We have therefore 
solved the pair conservation equation to estimate $-v/Hr$ which 
approaches unity on small scales, as required for consistency with
the results for $\xi$. 

\noindent $\bullet$ We find that the onset of stable clustering 
occurs at $x/x_{nl}(a) = 0.07$ for all spectra tested. 
This provides a useful way to demarcate the stable clustering
regime for generic spectra. The range  of $\xi$ over which
stable clustering is verified is typically $200<\xi<2000$; it  
is higher for initial spectra with more small scale power ($n\simeq 0$). 

\noindent $\bullet$ For the CDM spectrum we find the range of scales
for $0.2<a<1$ for which the evolution of $\xi$ is consistent with
stable clustering. The combination of simulations with two different 
box-sizes enabled us to study clustering on very small scales, 
typically down to comoving scales 
$=40-120 \, {\rm kpc}$ (with $h=0.5$), while retaining the nonlinear
influence of sufficiently long waves on the dynamics. The physical
scale demarcating the stable clustering regime rapidly increases
with time, and is about $r=20, 180, 700 \, {\rm kpc}$
for $a=0.2, 0.5, 1$ respectively. 

\noindent $\bullet$ The amplitude of the nonlinear $\xi$ agrees
to better than $20\%$ with the prediction of the Press-Schechter 
model of Sheth \& Jain (1996) for nearly all the spectra studied. 
For the CDM spectrum the results agree with the predicted slope and 
amplitude of the $n=-2$ spectrum. The agreement of the model for
different initial spectra is remarkable in view of the fact that
their nonlinear behavior is quite different: e.g. for $n=0$
the nonlinear $\xi$ is suppressed relative to the linear
$\xi$ by an order of magnitude, while for $n\simeq -2$ it is 
enhanced by about the same factor. 

\noindent $\bullet$ We emphasize that the commonly used parameterization 
$\xi=a^{3+\epsilon} \, (r/r_0)^{-1.8}$, with $\epsilon=0$ is accurate
only in the stable clustering regime of $r/r_{nl} < 0.07$.
It can severely underestimate
the growth of $\xi$ in the intermediate regime of 
$0.07< r/r_{nl} < 1$ in which the growth of $\xi$  
varies rapidly with $r$, ranging from $\epsilon\simeq 0-2$. The results
shown in Figures 6 and 7 can be used to estimate $\epsilon$ for
given $r$ and $a$ for CDM-like spectra.


For a numerical study aimed at testing an asymptotic regime
it is appropriate to conclude on a note of caution. 
Small scale clustering is closely linked to the structure of the 
inner parts of dark halos. Both the number density and density profile
of very small halos, with $M\ll M^*/10$, are poorly resolved
in N-body simulations -- this can crucially affect the shape
of $\xi$ in the small scale limit (S. White, private communication;
Sheth \& Jain 1996). 

Thus while the simulations used in this study provide strong evidence 
that gravitational clustering stabilizes on scales smaller than $\sim 1/10$th
the nonlinear scale, it is premature to draw conclusions about 
a possible asymptotic regime. We have shown results typically over a factor
of $5$ in length scale, and an order of magnitude in $\xi$, by 
using self-similar scaling to combine results from different output
times. The competing demands of small scale resolution and a
simulation box sufficiently bigger than the non-linear scale,
do not leave enough dynamic range to see a convincing asymptotic regime
on small scales (say a range of scales $\gg 10$). 

We have carefully analyzed the effects of
limited numerical resolution to rule out the possibility of a 
conspiracy of numerical artifacts over the range of scales that
were used in our analysis. Therefore we can conclude with confidence
that any departures from stable clustering could occur only on scales 
with $\xi\gg 10^4$, which lie beyond our resolution limits. 
We shall have to wait for the next generation of high resolution 
simulations, with $N \gsim 10^8$ particles, to find possible new aspects 
of nonlinear clustering. 

\section*{Acknowledgments}
I am very grateful to Ed Bertschinger and Simon White for
making available their simulation data for this paper, and
for stimulating discussions. I thank Martin Haehnelt, 
Uro\v s Seljak, Ravi Sheth and Bepi Tormen for helpful discussions.

\end{document}